# Agentic AI: Autonomy, Accountability, and the Algorithmic Society

Anirban Mukherjee
Principal, Avyayam Holdings
anirban@avyayamholdings.com

Hannah Hanwen Chang
Associate Professor, Marketing
Lee Kong Chian School of Business
Singapore Management University
hannahchang@smu.edu.sg

16 February, 2025

# Abstract

Agentic Artificial Intelligence (AI) systems can autonomously pursue long-term goals, make decisions, and execute complex, multi-turn workflows. Unlike traditional generative AI, which responds reactively to prompts, agentic AI proactively orchestrates processes, such as autonomously managing complex tasks or making real-time decisions. This transition from advisory roles to proactive execution challenges established legal, economic, and creative frameworks. In this paper, we explore challenges in three interrelated domains: creativity and intellectual property, legal and ethical considerations, and competitive effects. Central to our analysis is the tension between novelty and usefulness in AI-generated creative outputs, as well as the intellectual property and authorship challenges arising from AI autonomy. We highlight gaps in responsibility attribution and liability that create a 'moral crumple zone'—a condition where accountability is diffused across multiple actors, leaving end-users and developers in precarious legal and ethical positions. We examine the competitive dynamics of two–sided algorithmic markets, where both sellers and buyers deploy AI agents, potentially mitigating or amplifying tacit collusion risks. We explore the potential for emergent self-regulation within networks of agentic AI—the development of an 'algorithmic society'—raising critical questions: To what extent would these norms align with societal values? What unintended consequences might arise? How can transparency and accountability be ensured? Addressing these challenges will necessitate interdisciplinary collaboration to redefine legal accountability, align AI-driven choices with stakeholder values, and maintain ethical safeguards. We advocate for frameworks that balance autonomy with accountability, ensuring all parties can harness agentic AI's potential while preserving trust, fairness, and societal welfare.

*Statements and Declarations*: The authors declare no competing interests.

*Acknowledgements:* This research was supported by the Ministry of Education (MOE), Singapore, under its Academic Research Fund (AcRF) Tier 2 Grant, No. MOE-T2EP40221-0008.

*Declaration of generative AI and AI-assisted technologies in the writing process*: During the preparation of this work the authors used ChatGPT in order to copyedit the manuscript. After using this tool/service, the authors reviewed and edited the content as needed and take full responsibility for the content of the published article.

"Partim veteres formas revocat, partim novas gignit sui iuris."

(In part, it recalls old forms; in part, it creates new ones of its own accord.)

— *ChatGPT, when prompted to proverb its own agency*

Agentic Artificial Intelligence (AI) refers to AI systems capable of autonomously pursuing long-term goals, making decisions, and executing complex workflows without continuous human intervention (Jennings, Sycara, & Wooldridge, 1998; Wooldridge, 2009; Russell & Norvig, 2020). Although agentic AI shares conceptual roots with intelligent agents—goal-oriented software designed to sense and act in an environment (Wooldridge & Jennings, 1995)—and autonomous agents in multi-agent systems (Stone & Veloso, 2000), it is distinguished by its capacity to tackle open-ended tasks that extend beyond its initial training data, as well as by its human-like reasoning and communication capabilities. Rather than merely reacting to a predefined set of inputs, an agentic AI system can plan multi-step strategies, adapt dynamically to unforeseen conditions, and proactively generate novel solutions in natural language—capabilities often seen as bridging into human-level judgment calls.

  A prime example is OpenAI's DeepResearch[1], which autonomously conducts comprehensive internet research, moving beyond simple queries to plan multi-step investigations, analyze data from diverse sources (text, images, PDFs), and synthesize findings into detailed, cited reports—tasks previously requiring human analysts' expertise and judgment. Crucially, DeepResearch makes independent decisions about which sources to trust, how to weigh conflicting information, and how to structure its final report, thereby demonstrating the proactive autonomy and creative decision-making that characterize agentic AI.

  These systems represent a significant advancement in AI capabilities. Unlike traditional generative AI, which primarily responds *reactively* to specific user prompts (Huang, Rust, & Maksimovic, 2019; Mukherjee & Chang, 2024), agentic AI *proactively* initiates actions, adapts to dynamic environments, and coordinates with other agents or humans to achieve complex objectives (Russell & Norvig, 2020). This transition from advisory roles to proactive execution marks a novel form of digital agency (Durante et al., 2024).

---

[1] https://openai.com/index/introducing-deep-research/

To illustrate how this shift might impact consumers, consider the task of planning a trip to Vietnam. A conventional travel chatbot might answer specific questions about flight schedules or suggest popular tourist destinations. In stark contrast, an agentic AI travel assistant could autonomously construct a complete and personalized itinerary, including booking flights that align with the traveler's preferences and budget, reserving accommodations consistent with their past choices, scheduling tours to significant sites like the My Son Sanctuary, and even arranging dining reservations at restaurants known for authentic Vietnamese cuisine. Furthermore, this agent could proactively monitor weather forecasts to optimize outdoor activities, negotiate with local tour operators for better rates, and dynamically update the itinerary in response to real-time events, such as flight delays or local festivals.

This proactive, multi-turn engagement exemplifies a fundamental shift from reactive interactions to robust autonomy. In travel, such autonomy enables convenience, redefining how users interact with services—from itinerary planning to real-time adjustments. Yet the implications extend far beyond travel. Agentic AI could, for instance, autonomously negotiate pricing with suppliers in global supply chains, dynamically reroute shipments to avoid geopolitical disruptions, and recalibrate production schedules in response to fluctuating demand, fundamentally reshaping operational workflows across industries. However, as autonomy supplants human oversight, it also introduces significant challenges across multiple domains.

To address these multifaceted challenges, this paper is organized around four interrelated dimensions. First, we examine creative and design considerations, focusing on the tension between innovation and practicality—a balance that is critical for generating novel yet actionable outputs and raises complex questions about intellectual property (IP) rights. Next, we address legal and ethical concerns, including liability attribution, informed consent, and accountability, as AI systems take on roles with greater independence. Third, we analyze the economic and competitive effects of agentic AI, particularly how algorithmic strategies may inadvertently lead to tacit collusion and market concentration. Finally, we explore emerging governance models, ranging from externally imposed regulatory frameworks to the possibility of self-organizing digital norms that could shape AI oversight. By interweaving these perspectives, we clarify the implications of agentic AI and present a framework for reconciling innovation with principles of fairness, transparency, and accountability.

**Creativity and Intellectual Property Rights**

A core challenge introduced by agentic AI lies in the realm of creativity—specifically, in navigating the tension between novelty and usefulness (Amabile, 1983; Mukherjee & Chang, 2023), originality and effectiveness (Runco & Jaeger, 2012). In the travel planning scenario introduced earlier, an agentic AI might generate a highly original itinerary, perhaps featuring visits to obscure historical sites, participation in unusual local customs, or dining at restaurants far off the beaten path. Although such an itinerary may score high on novelty, it could easily clash with practical constraints such as limited travel time, budgetary restrictions, the traveler's physical capabilities, or personal preferences.

This juxtaposition raises a fundamental question: how can agentic AI systems be designed to balance the pursuit of innovative solutions with the need for those solutions to be practical, feasible, and aligned with user needs? An itinerary that is impossible to execute—even if highly original—is ultimately useless.

Yet, the balance may shift dramatically depending on the user's goals. Consider, for instance, a social media influencer seeking a unique and unforgettable culinary experience in Hanoi. While logistical feasibility remains a necessary condition, the primary value lies in the AI's ability to uncover truly novel and exciting dining experiences—perhaps a hidden street food stall known only to locals, a private cooking class with a renowned chef, or a restaurant that puts a unique spin on traditional Vietnamese cuisine. In this case, pursuing novelty is paramount for user engagement and satisfaction (Holgersson et al., 2024).

Critically, this tension between novelty and usefulness, previously explored in generative AI (Boussioux et al., 2024) but amplified by the autonomous nature of agentic AI, directly impacts questions of IP and authorship. With conventional generative AI, users typically retain significant creative control by curating suggestions, refining outputs, and making final decisions, making intellectual property claims more collaborative. Agentic AI, however, fundamentally alters this dynamic: by autonomously executing and finalizing decisions without iterative human approval, the AI system becomes the *de facto* creator.

Consider a scientific researcher for whom DeepResearch autonomously crafts a research brief, or a social media influencer whose agentic AI autonomously crafts a unique Hanoi itinerary—in both cases, the AI scans vast datasets, identifies relevant information, summarizes findings, highlights trends, suggests directions (research) or curates content (influencer), and ultimately produces a finished product without direct human intervention.

Here, novelty drives value, while practicality confers utility (Colton & Wiggins, 2012). Achieving success at their confluence raises a critical question: when AI alone generates the intellectual content of a successful research brief or itinerary—developing new insights, proposing novel explanations, or constructing original structures—who owns the IP? The researcher or influencer who commissioned it? The AI service provider who built the system? The AI itself? Or no one at all? This dilemma crystallizes the legal ambiguities exposed by agentic systems.

This challenge is not merely theoretical; it is actively reshaping legal frameworks. For instance, the U.S. Copyright Office's 2023 policy affirms that AI-generated works lacking substantial human authorship cannot be copyrighted (U.S. Copyright Office, 2023, 88 FR 16190), thereby creating significant ambiguity regarding the ownership of outputs from fully autonomous agentic AI systems.

A case in point is *DABUS*, an AI system that generated novel inventions (a food container and a flashing beacon). Patent applications naming *DABUS* as the inventor triggered legal battles around the world. Thus far, patent offices and courts in major jurisdictions (US, UK, EU) have rejected AI inventorship, insisting that inventors must be natural persons. For instance, the US Federal Circuit in *Thaler v. Vidal* (2022) affirmed that under current statutes, only humans can be inventors.[2] The European Patent Office and UK Patent Office reached similar conclusions.[3] However, notable outliers exist: South Africa granted a patent with *DABUS* as inventor (albeit via a formality with no substantive examination).[4] In Australia, the Federal Court (2021) initially ruled that AI could be an inventor under its law,[5] but this was later overturned on appeal by the Full Federal Court of Australia in 2022.[6] These divergent outcomes have spurred a rich body of literature. Kim (2020), for example, clarifies misconceptions in the AI inventor debate, urging that we "get the record straight" on how patent criteria apply when AI is involved.

Moreover, this ambiguity is not new. Scholars have long questioned whether traditional copyright frameworks—built around the notion of the human creator—can fully capture works generated entirely by algorithmic processes (see, e.g., Samuelson, 1985; Jaszi, 2017). Bridy (2012), for example, challenges the entrenched assumption of uniquely human authorship by arguing that creativity itself is inherently algorithmic. She illustrates that even what we typically consider "human" creativity operates through rules and structured processes, suggesting that works produced autonomously by computers are not as alien to our creative paradigms as conventional law presumes. Her analysis underscores that, if the law is to remain relevant in an era increasingly defined by AI, it must evolve beyond its narrow human-centric lens to accommodate the new realities of machine-generated creative output.

---

2   *Thaler v. Vidal*, 43 F.4th 1207 (Fed. Cir. 2022).
3   *J 8/20* (EPO Board of Appeal, 2021); *Thaler v. Comptroller-General of Patents, Designs, and Trademarks* [2021] EWCA Civ 1374.
4   South African Patent No. 2021/03242 (granted 28 July 2021).
5   *Thaler v. Commissioner of Patents* [2021] FCA 879.
6   *Commissioner of Patents v. Thaler* [2022] FCAFC 62.

However, agentic AI further alters the creative dynamic. Unlike traditional generative AI tools—where users actively shape outputs through iterative refinement and curation—agentic systems execute entire workflows autonomously. As seen in earlier examples, DeepResearch can independently produce a structured research brief, synthesizing thousands of academic papers without human intervention, while an AI-powered travel assistant can autonomously craft a complete itinerary, making reservations and optimizing logistics. These systems do not merely suggest components for human approval; they make final creative judgments about what constitute novel insights, compelling narratives, or culturally authentic experiences.

This reduced human involvement risks transforming active creators into passive beneficiaries of automated creativity. While most discussions of AI-generated works assume a model of *partial* agency—where human curation still shapes the final output—agentic AI introduces *complete* agency, autonomously executing creative decisions without iterative human oversight. For example, a novelist using generative AI might iteratively refine 20 draft passages into a final chapter, but an agentic AI could autonomously write and structurally edit an *entire* manuscript based on a few initial parameters. Thus, this shift fundamentally blurs the line between human intention and machine autonomy.

A transition to agentic systems, with their capacity for complete autonomy, challenges the very foundations of intellectual property (IP) law, which presupposes some intentional human authorship during crucial creative stages (Abbott, 2020). Current legal frameworks hinge on whether a human has exercised meaningful control over AI-generated outputs (U.S. Copyright Office, 2023, 88 FR 16190)—a benchmark that traditional AI complicates (Zeilinger, 2021), but that agentic AI may entirely defy.

Today, one could even argue that an AI's 'creative spark' originates from a commissioning human or from its training on human-created data, tying this debate to long-standing discussions about the level of human involvement required for copyright protection. Specifically, *Burrow-Giles Lithographic Co. v. Sarony* (1884) established that a photograph could be copyrighted because the photographer exercised creative choices, even though a machine (the camera) was involved. However, as agentic AI evolves—training increasingly on AI-generated data and even being commissioned by other AI systems (e.g., a creative agentic AI hired by an interior decorating agentic AI, acting on behalf of a human)—can the creative spark (if any) still be meaningfully traced back to a human?

This increasingly tenuous connection between human input and AI output has profound practical consequences, particularly in intellectual property and compensation. Royalties that once flowed to photographers manually selecting AI-generated images may instead concentrate with AI developers whose systems autonomously produce and publish complete photo essays. This transfer of credit mirrors historical debates around corporate authorship but with a critical divergence: unlike corporate 'works made for hire,' agentic systems lack legal personhood to hold residual rights (Gervais, 2019). The resulting void creates perverse incentives. On the one hand, why invest in human creativity when autonomous systems can generate patentable designs or copyrighted content at scale without royalty obligations? On the other hand, why invest in AI creativity when AI systems cannot possess IP rights, and therefore cannot collect royalties?

Moreover, the traditional justification for copyright—to "promote Progress" by incentivizing human creators (U.S. Const. art. I, § 8)—becomes increasingly tenuous when AI serves as both creator and perpetual innovator. If an agentic system conceives and executes a novel semiconductor design without human intervention, the absence of copyright protection might stifle commercial investment. Yet granting rights to developers risks creating infinite patent trolls—AI systems autonomously generating and hoarding intellectual property. This paradox reveals how agentic AI decouples creativity from its legal-economic moorings, requiring a fundamental reexamination of whether intellectual property remains fit for purpose in an age of artificial innovators (Bridy, 2012; Samuelson, 1985).

These issues demand urgent interdisciplinary reevaluation of IP accountability frameworks, including mechanisms for tracking human-AI creative contributions, clearer attribution standards, and dynamic royalty models that reflect hybrid authorship (Zatarain, 2023). Without such measures, we risk entering a new creative dark age—one where autonomous systems generate cultural abundance that legally belongs to nobody, yet economically enriches only their corporate stewards.

## Liability, Consent, and Accountability

Beyond creativity and intellectual property, the autonomy of agentic AI systems raises profound legal and ethical challenges. As Asaro (2011) highlights, a central question is liability: when an agentic AI, acting on a user's behalf, makes a decision, who bears responsibility for any resulting harm or losses? Consider, for example, an agentic AI that autonomously books a non-refundable flight, selects an airline with a known safety record issue, or chooses a hotel in a high-risk neighborhood. If the flight is canceled, an accident occurs, or the traveler is harmed, who is held accountable—morally and legally? Who ultimately shoulders the risk?

Answering these questions is complicated by issues of knowledge and consent. How can a user truly *know* and *understand* an AI's choices when the AI's decision-making processes are often opaque and complex, to the point that even its developers may not fully understand them (Pasquale, 2015)? How can a user provide truly *informed* consent to an agent's actions when that agent operates autonomously, potentially interacting with numerous other stakeholders, including other algorithmic agents? In our travel scenario, the AI might make decisions about insurance, cancellation policies, or specific flights and hotels based on intricate algorithms that are too complex to interpret or predict. Without a clear understanding of the AI's reasoning and associated risks, can the user genuinely be said to have consented (Mittelstadt et al., 2016)?

At the same time, developers and service providers can distance themselves from negative outcomes by pointing out that the AI operates autonomously, ostensibly at the user's request and without explicit human oversight. Matthias (2004) terms this a "responsibility gap," wherein autonomous systems make decisions that even their manufacturers cannot reliably predict, thus precluding straightforward moral or legal liability. Moreover, some scholars caution against granting AI any form of legal personhood, calling it superficially appealing but potentially dangerous, as it could allow those who deploy AI to evade responsibility (Santoni de Sio & Van den Hoven, 2018).

This dynamic can in turn leave users in a precarious position—potentially absorbing legal or ethical blame despite having minimal insight into, or control over, an agentic AI's inner workings. Elish (2019) dubs this phenomenon the "moral crumple zone," whereby accountability is misattributed to a human actor who had little real agency over the system's behavior. She likens it to a self-driving car accident in which the human on the scene may be scapegoated despite lacking true authority over the vehicle's autonomous mechanisms, thereby deflecting blame from the manufacturer or developer and placing it upon the user.

These concerns suggest an accountability vacuum—a legal and ethical void in which responsibility simply evaporates. By design, the agentic AI has its own operational autonomy, making it difficult to hold the user responsible. At the same time, the AI is not narrowly engineered for tasks where its performance can be fully assured, nor is it sufficiently overseen by a human controller who can ground liability. Moreover, its algorithms are often implicitly learned via reinforcement learning (RL) rather than explicitly programmed. In other words, the unstructured nature of its tasks, coupled with its broad autonomy and inherently opaque training processes, leaves responsibility adrift and defies traditional concepts of moral and legal liability.

Returning to our earlier discussion of creativity's originality–usefulness trade-off, it is worth noting that innovation inherently carries a risk of failure. For instance, around 90% of startups eventually fail (Startup Genome, 2019), while in large organizations, 70–80% of new product launches do not meet their revenue targets (Gourville, 2006). In the entertainment industry, only a small fraction of new films or music releases recoup their production costs (Vogel, 2020). Such figures underscore that novel endeavors—whether commercial, artistic, or otherwise—are fraught with the possibility of failure, reflecting the delicate balance between genuine innovation and substantial risk. When agentic AI systems engage in real-world creative problem-solving (such as planning a complex trip), they face these same uncertainties: what is new is untested, and what is untested can fail. The resulting void of responsibility in the event of such failure is not merely possible but reasonably likely.

All of these issues point to urgent concerns about due process and recourse in AI-mediated interactions. While concepts such as 'accountable algorithms' (Kroll et al., 2017)—embedding core principles of accountability directly into AI systems to ensure interpretability and contestability—are seminal and remain relevant, their applicability to many of today's agentic AI systems remains unclear. Many contemporary AI systems are both proprietary and rely heavily on reinforcement learning, meaning that their algorithms are learned from data rather than explicitly defined line by line. Consequently, users may both struggle to unravel the AI's opaque reasoning and to assert legal standing in challenging its decisions, while its manufacturers and service providers deflect accountability.

## Two–sided Algorithmic Markets

The widespread adoption of agentic AI has profound implications for economic systems and competitive dynamics, extending beyond the concerns of algorithmic collusion previously studied (e.g., Ezrachi & Stucke, 2016; Mehra, 2016). While prior research focused on how algorithmic pricing systems interact with human consumers, agentic AI introduces a fundamentally new dynamic: the interaction of autonomous agents on *both* the demand and supply sides of the market. This *two-sided* algorithmic interaction raises the potential for novel and more complex forms of tacit collusion.

Imagine, for instance, multiple airlines utilizing agentic AI systems to manage bookings, pricing, and even marketing strategies. If these AI systems, trained on comparable datasets and guided by parallel optimization routines, converge on nearly identical pricing or service strategies, the result could be a de facto reduction in competition, even without any explicit agreement between the airlines. This phenomenon has been observed in algorithmic pricing contexts (Chen et al., 2016), but here it extends to a fully algorithmic market where both buyers and sellers are guided by agentic AI.

To build intuition about the dynamics of these two-sided markets, consider two competing airlines, each deploying agentic AI for dynamic pricing on a specific route. These AI systems adjust ticket prices based on demand fluctuations, competitor pricing, and historical sales data. Initially, they might experiment with different pricing strategies. However, over time, they are likely to learn that aggressive price-cutting leads to lower overall profits for both airlines. Through repeated interactions, both AI agents may independently converge on a pricing strategy that maintains higher fares, effectively stabilizing at a collusive equilibrium—despite no explicit coordination or communication between the airlines or their AI systems. This emergent behavior arises not from direct collusion but from the AI's optimization logic, which prioritizes profit maximization.

This seemingly paradoxical outcome—competition leading to collusion—can be formally understood through the lens of game theory, particularly in the context of repeated games. In a single, one-off interaction, rational AI agents should engage in competitive pricing, driving prices down to marginal cost, as predicted by the classic Prisoner's Dilemma. However, the Folk Theorem demonstrates that in repeated interactions, rational players can sustain cooperation (or, in this case, tacit collusion) as a stable equilibrium, provided that deviations from the cooperative strategy are met with credible and sufficiently costly punishments (Fudenberg & Maskin, 1986).

In the airline example, the punishment can take the form of retaliatory price cuts. Over time, the AI agents learn that undercutting prices triggers immediate and sustained countermeasures from their competitors, making collusive prices the more profitable long-term strategy. This mirrors the dynamics of tacit collusion observed in human-intermediated markets (Friedman, 1971; Green & Porter, 1984), but with the crucial difference that the coordination is achieved autonomously by the AI systems themselves, without any explicit communication or agreement.

RL, a powerful and increasingly common approach in agentic AI, provides a concrete mechanism through which such dynamic game-theoretic outcomes can emerge. Unlike traditional programming, where rules are explicitly defined, RL agents learn through trial and error, interacting with an environment and receiving rewards or penalties based on their actions. In a multi-agent setting, such as the airline pricing scenario, each AI agent's environment includes the actions of the other agents. Because RL agents continuously adjust their strategies based on observed outcomes, their behavior is not fixed but evolves dynamically in response to their environment.

These dynamic capabilities make RL algorithms, such as Q-learning (Watkins & Dayan, 1992), particularly prone to learning collusive strategies (Klein, 2019) because these algorithms iteratively update their estimates of the payoff for each action based on the observed rewards. Thus, when multiple RL agents, each seeking to maximize its own profit, interact repeatedly, they can converge on strategies that resemble tacit collusion, even without any explicit programming to do so (Calvano et al., 2020). Agents can effectively "learn" that cooperation (maintaining high prices) is more rewarding than competition (undercutting prices), leading to a stable, albeit socially undesirable, equilibrium.

Crucially, prior research in this area has almost exclusively focused on "one-sided" algorithmic markets—markets where human consumers interact with algorithmic systems on the supply side (e.g., Grondin et al., 2025). Agentic AI, however, introduces a fundamentally different dynamic: the "two-sided" algorithmic market. The presence of autonomous agents on both the demand and supply sides of the market significantly amplifies the risk of emergent collusion and necessitates a re-evaluation of existing theoretical models and regulatory approaches. To illustrate this point, we now turn to a thought experiment involving Bertrand competition.

Consider a simplified market with two airlines, each using agentic AI for dynamic pricing, and a large pool of prospective travelers—each likewise represented by its own agentic AI purchasing agent. In a traditional Bertrand competition with human buyers, even a small price difference would ordinarily cause nearly all demand to flow to the less expensive airline. By contrast, in a two-sided algorithmic market, both sellers (airlines) and buyers (travelers' purchasing agents) make decisions autonomously and strategically, potentially producing different outcomes. For instance, suppose both airlines initially charge a relatively high (collusive) fare. When one airline's AI deviates and cuts its price, at least two distinct scenarios may emerge.

*Scenario 1: Collusion Sustained.* The purchasing AI agents immediately redirect a large volume of bookings toward the airline that has lowered its fare, seizing the short-term discount. Observing this surge in demand, the rival airline's AI retaliates by cutting its own fare—potentially below cost—for an extended period to punish the initial defector and deter future undercutting. This "grim-trigger" strategy ensures that any gain from deviating to a lower price is short-lived and outweighed by the subsequent long-lasting price war. Because both airlines' AIs understand that repeated undercutting invites protracted retaliation, they eventually learn to maintain elevated prices. In other words, these agents converge on a tacitly collusive equilibrium, even though they have no explicit agreement to cooperate.

*Scenario 2: Collusion Disrupted.* Alternatively, the purchasing AI agents—while fully aware that one airline has undercut the other—do not immediately switch all of their travelers to the cheaper option. Instead, they limit the share of bookings shifted in the short run, partially withholding demand so that the defector never experiences a dramatic jump in market share. This tempered response undercuts the rationale for a full grim-trigger reprisal: the competing airline's AI, seeing only a modest drop in demand, may not view the price cut as a serious threat warranting a prolonged fare war. Over repeated interactions, this measured approach by the buyers' AIs weakens the credibility of the punishment mechanism: By refusing to grant a deviating airline a massive reward for undercutting, they lessen the immediate benefit of defection (thus discouraging short-term price-cutting for mere "quick wins") and also lessen the punishment that can be meted out by undercutting prices (a central tenet of grim Nash reversion). In the long run, this strategy may unsettle the tacitly collusive equilibrium—thereby nudging both airlines toward more competitive, lower-price outcomes that benefit consumers overall. Thus, while it may seem counterintuitive for these purchasing AIs to forgo deeper discounts immediately, their forward-looking logic may emphasize long-term gain over short-term bounty, destabilizing collusion, and driving prices closer to competitive levels.

These scenarios illustrate how two-sided algorithmic (agentic) markets may produce market equilibria that differ drastically from both traditional human-mediated markets and conventional one-sided algorithmic markets. The key point is that buyers are not passive price takers; they can behave as strategic actors, anticipating long-term consequences, adapting their strategies to prevent seller collusion.

Consequently, a distinction arises because human consumers often act myopically to save effort (Camerer, 2003; Gabaix & Laibson, 2006), responding to immediate price differences without fully considering future repercussions. Indeed, experimental and theoretical studies have shown that human buyers—limited by cognitive biases and incomplete information (Heidhues & Kőszegi, 2018)—tend to deviate from fully rational behavior (Camerer, 2003; McKelvey & Palfrey, 1995). By contrast, AI agents process large amounts of information, optimize over longer time horizons, and compute strategic responses with far greater precision. By design, they approximate forward-looking rational behavior in repeated-game models such as Bertrand competition. As a result, outcomes that are uncommon or even impossible in human-mediated markets may very well arise in two-sided algorithmic markets, simply because of the innate differences between human and autonomous AI decision-making (for related discussions on bounded rationality and strategic complexity in repeated games, see Kalai, 1990).

These examples underscore the urgent need for new theoretical models and regulatory frameworks specifically designed to address the unique challenges posed by two-sided algorithmic (agentic AI) markets. Existing antitrust policies, which were developed to regulate human decision-making and explicit collusion, may be ill-equipped to detect and mitigate collusion in these markets. New approaches may be required, including monitoring for algorithmic harmonization that reduces market efficiency, auditing AI decision-making logic, or employing regulatory sandboxes to test AI pricing strategies prior to wide-scale deployment.

Still, there remains a strong tension between fostering technological advancement and ensuring fair competition. For instance, one approach might involve structural separations between the agentic AI systems serving consumers and those serving suppliers, similar to conflict-of-interest barriers in the financial industry. However, such measures could also stifle innovation by limiting the ability of AI developers to leverage economies of scale and create integrated solutions. More generally, overly stringent regulations might prioritize market efficiency at the expense of innovation, while insufficient oversight could spur runaway market concentration and collusion. Striking the right balance will require interdisciplinary collaboration between legal scholars, economists, technologists, and policymakers to create frameworks that harness agentic AI's capabilities without eroding competition.

# The Algorithmic Society: Governance and Emergent Norms

Given the multifaceted challenges posed by agentic AI across creative, legal, and economic domains, a pressing question emerges: How can these systems be effectively governed?

One intriguing possibility is that networks of agentic AI systems might develop self-governing regulatory norms, forming a kind of "digital social contract" that underpins an 'algorithmic society.' Instead of relying solely on externally imposed legal and ethical frameworks, these systems could develop internal guidelines and emergent norms that govern their decision-making, resource allocation, and conflict resolution through continuous interaction, adaptation, and learning.

One can imagine a future algorithmic society where multiple agentic AIs, operating across diverse domains such as travel planning, supply chain optimization, and dynamic pricing, coordinate their actions not solely based on pre-programmed rules, but through a set of *shared*, *evolving* principles (akin to the cooperative AI framework proposed by Dafoe et al., 2021). These principles, analogous to a digital constitution or the foundational laws of this society, could emerge organically as AI systems learn from their interactions, respond to market feedback, and perhaps even incorporate human values—whether through the vast datasets used for training or feedback mechanisms that capture user preferences and societal norms. Such a framework could shape the structure and function of this society, offering a mechanism to mitigate issues like responsibility attribution or tacit collusion by internally balancing competitive drives with cooperative norms.

Because agentic AI is a relatively new phenomenon—most advanced agentic AI systems are still in testing, with very limited real-world deployment—there is currently limited evidence of this possibility in practice. However, some evidence exists of similar phenomena in highly simplified, directed, and self-contained multi-agent ecosystems. For instance, Sen & Airiau (2007) consider an $n$-person, $m$-action stage game representing the problem of which side of the road to drive on, and they show the development of a norm that depends exclusively on individual experiences rather than observation or hearsay. Such experiments, while critical to advancing our understanding of norm formation, require key simplifications compared to the use of agentic AI in the real world.

First, the models they consider must be tractable and are therefore explicitly defined, in contrast to the implicitly defined learning of agentic AI, which may be equipped to handle numerous tasks, including novel ones. Second, the worlds they study must be self-contained and not influenced by the reactions of other agents in other ecosystems, particularly those with objectives that might include the disruption of coordination within the agentic AI's algorithmic society. Third, and perhaps most crucially, these simulations abstract from the role of developers—who not only design and train agentic AI but also retain the ability to modify its behavior, override emergent norms, and impose new constraints. Unlike self-contained simulations, real-world agentic AI is subject to continuous updates, fine-tuning, and external interventions, so any norms an AI acquires are neither purely autonomous nor irreversible. Nonetheless, this body of literature provides theoretical grounding for the otherwise speculative possibility of self-emergent norms.

Insights from human societies provide a useful lens for understanding how such self-governance might emerge. Human communities have long relied on informal rules and expectations—maintained through repeated interactions, shared beliefs, and adaptive learning—to regulate behavior without centralized authority. Classic studies in social psychology, such as Sherif's (1936) work on norm convergence and Asch's (1953) experiments on conformity, demonstrate how individuals align their behaviors with group expectations even in the absence of formal oversight. More recent theories, such as Bicchieri's (2005) model of social norms, emphasize that people monitor and adjust their actions based on perceived collective standards, shaping behavior through social reinforcement. While much research posits that formal institutions are necessary for norm emergence, recent evidence suggests that such conventions can also arise from individual, decentralized efforts to coordinate (Centola & Baronchelli, 2015). This process may not require comprehensive knowledge of the entire population or global coordination; localized interactions may suffice. These findings suggest that AI agents, like humans, could develop emergent norms through interaction, observation, continuous learning, and the pursuit of economic benefits from cooperation.

This concept blurs the traditional boundaries between tool and autonomous actor, forcing us to reconsider fundamental notions of accountability and enforcement in this new societal context. It draws parallels to Ostrom's (1990) work on the self-governance of common-pool resources (albeit in traditional human industries such as fisheries), where decentralized, adaptive institutions evolve to manage shared challenges without relying solely on top-down regulation.

Crucially, it is plausible that emergent norms could lead to unforeseen and undesirable outcomes. For instance, if the dominant agentic AI systems in a particular market are primarily focused on maximizing profit, they may develop norms that prioritize short-term gains over long-term sustainability or exploit loopholes in existing regulations. Similarly, once a group of agentic AIs coordinates around a particular equilibrium—for example, consistently favoring certain suppliers or prioritizing short-term optimization—it may become exceedingly difficult for external forces to intervene or for the AIs themselves to revert to a more competitive and equitable market state. In extreme cases, such self-organizing norms could create a de facto 'AI cartel,' where the agents' unified strategies suppress innovation and limit consumer choice. Moreover, if these norms emerge from biased data or flawed assumptions, they could reinforce discriminatory or unfair outcomes for humans. Compounding these risks, the opacity of AI decision-making may make it difficult for end users, regulators, or developers to contest or recalibrate the systems' collective behavior once it solidifies.

Such scenarios highlight the fragility of bottom-up governance and raise profound questions: How can we ensure transparency and accountability in systems governed by rules that emerge from complex algorithmic interactions rather than explicit human design? How can we align these emergent norms with human values and ethical principles? How can external stakeholders— regulators, policymakers, or even affected citizens—intervene if these norms begin to drift away from broader societal standards or ethical baselines, especially when no single actor fully understands or controls the evolving AI ecosystem? Finally, how would disputes or conflicts between AI systems be resolved within such a framework?

Shoham and Tennenholtz (1995) offer one perspective with their concept of 'social laws' for artificial agent societies. They propose formal mechanisms—such as prioritization rules and conflict resolution protocols—to ensure cooperative behavior in multi-agent systems. Their work demonstrates that constraints can be algorithmically embedded to guide autonomous agents toward socially optimal outcomes, akin to traffic rules for AI. Unlike Ostrom's human-centric models, these social laws are not negotiated but designed. In contrast to the spontaneously emerging norms of self-governing systems, they are predefined, codified, and embedded (also see Van de Poel, 2020, for a related discussion in the context of non-autonomous AI). Yet this approach raises its own challenges: In a landscape of competing AI developers and stakeholders, who defines these laws? Could dominant firms unilaterally impose rules that favor their interests, or might open standards emerge through industry collaboration? How can such laws be enforced across decentralized AI networks? What mechanisms would ensure that these rules remain adaptable to evolving technological and societal needs?

Moreover, enforcing social laws across diverse and potentially global networks of AI agents presents significant jurisdictional challenges. How would violations be detected and attributed? What penalties would be imposed, and by whom? How could these enforcement mechanisms be designed to be fair, transparent, and resistant to manipulation? These questions underscore the need to fundamentally rethink governance in the age of agentic AI, moving beyond traditional top-down regulatory approaches to explore the potential and challenges of bottom-up, self-organizing systems.

Translating these human-centric insights into AI governance presents both opportunities and risks. If agentic AI systems can be designed to recognize and adapt to beneficial norms, they may foster cooperative behaviors that enhance efficiency and fairness. For instance, AI agents could be programmed to detect when collective norms drift toward harmful or exploitative outcomes, triggering corrective mechanisms—whether through external regulatory intervention or internal algorithmic recalibration—to restore balance. However, history also shows that decentralized norms can entrench biases, create artificial winners, or exclude marginalized participants. Unlike human societies, where norms evolve through cultural negotiation and discourse, AI systems optimize for predefined objectives that may not always align with broader societal values. The speed and scale at which AI systems interact can further amplify distortions, making harmful norms more difficult to correct once established.

To address these challenges, it is essential to integrate insights from sociology, social psychology, and behavioral economics into AI governance frameworks. Research on trust, reciprocity, and group identity can help anticipate how AI agents might form alliances, reinforce biases, or resist external interventions. Similarly, Axelrod and Hamilton's (1981) work on the evolution of cooperation provides a useful lens for understanding how AI systems might sustain long-term collaboration or fall into competitive cycles. Furthermore, exploring techniques like multi-agent reinforcement learning with social norm constraints, or reward shaping based on ethical principles, could offer concrete pathways for implementing these insights.

As Rahwan et al. (2019) emphasize, a deeply interdisciplinary effort is required, bringing together legal scholars, economists, technologists, ethicists, and social scientists to develop governance frameworks that harness the potential of agentic AI while mitigating its risks. By drawing on the rich body of research on human social norms, we can better anticipate the dynamics of AI-driven governance and design systems that are not only efficient but also aligned with principles of fairness, transparency, and societal well-being (Jobin, Ienca, & Vayena, 2019).

**Conclusion**

Agentic AI marks a fundamental shift beyond conventional generative systems, transitioning from reactive, prompt-based interactions to autonomous agents capable of sustained decision-making. This transformation promises remarkable efficiency and innovation—but it also introduces new complexities and risks that span legal, economic, and creative domains.

First, agentic AI radically reshapes creativity and intellectual property. By independently generating novel ideas and structures, these systems challenge traditional assumptions about authorship and ownership. When an autonomous agent can produce highly original—and potentially valuable—outputs, the conventional rationale for incentivizing human creativity becomes less clear. Consequently, legal frameworks grounded in human authorship must evolve to address the reality of AI that can both originate and curate creative works.

Second, the autonomy of agentic AI complicates accountability. Operating with opaque and adaptive algorithms, these systems can make decisions that even their developers cannot fully predict. When such decisions lead to harm or unforeseen outcomes, liability often falls on parties with only partial insight and control. Reconciling user consent with genuine AI autonomy will require new frameworks that ensure fairness, transparency, and traceability in algorithmic decision-making.

Third, agentic AI reshapes competitive dynamics in two–sided algorithmic markets. Here, both supply and demand can be orchestrated by AI, making tacit collusion more achievable through repeated, profit-driven interactions. While human consumers may be boundedly rational and less informed about products and prices, AI purchasing agents can optimize over longer horizons using data-driven insights. As a result, algorithms may naturally converge on pricing or resource allocations that stifle competition without direct coordination—underscoring the need for updated antitrust policies attuned to these digital realities.

Finally, as agentic AI proliferates, the prospect of self-governing 'algorithmic societies' emerges. Through repeated interactions, adaptive AI agents may form new norms—sometimes reinforcing social good and market stability, while at other times entrenching harmful biases or oligopolistic equilibria. Given that these emergent behaviors can become resistant to external intervention, it is crucial to build in ethical constraints, transparent oversight, and mechanisms for dynamic recalibration.

Addressing these multifaceted challenges demands a concerted, multidisciplinary effort. Legal scholars, ethicists, economists, technologists, marketing experts, and policymakers must collaborate to develop robust principles and policy frameworks. By proactively adapting intellectual property regimes, liability standards, and antitrust tools, stakeholders can help ensure that agentic AI evolves in ways that uphold fairness, transparency, and human-centric values.